\documentstyle[twoside,epsf,11pt]{article}
\newcommand {\oalf} {\mbox{${\cal O}(\alpha)$}}

\newcommand{\epl}{$e^+$}
\newcommand{\emi}{$e^-$}
\newcommand{\ee}{$e^+ e^-$}
\newcommand{\nl}{\nonumber \\}
\newcommand{\bq}{\begin{equation}}
\newcommand{\eq}{\end{equation}}
\newcommand{\ba}{\begin{eqnarray}}
\newcommand{\ea}{\end{eqnarray}}
\newcommand{\ww}{$W^+ W^-$}
\newcommand{\zz}{$Z^0$}

\newcommand{\ltwo}{LEP~2~}
\newcommand{\RS}{\sqrt{s}}

\newcommand{\EE}{e^+ e^-}
\newcommand{\ZZ}{Z^0}

\newcommand{\para}{\par\noindent}
\def\photonatomright{\begin{picture}(3,1.5)(0,0)
                                \put(0,-0.75){\tencircw \symbol{2}}
                                \put(1.5,-0.75){\tencircw \symbol{1}}
                                \put(1.5,0.75){\tencircw \symbol{3}}
                                \put(3,0.75){\tencircw \symbol{0}}
                      \end{picture}
                     }

\def\photonatomup{\begin{picture}(1.5,3)(0,0)
                             \put(-0.75,3){\tencircw \symbol{3}}
                             \put(-0.75,1.5){\tencircw \symbol{2}}
                             \put(0.75,1.5){\tencircw \symbol{0}}
                             \put(0.75,0){\tencircw \symbol{1}}
                   \end{picture}
                  }

\def\photonright{\begin{picture}(30,1.5)(0,0)
                     \multiput(0,0)(3,0){10}{\photonatomright}
                  \end{picture}
                 }

\def\photonrighthalf{\begin{picture}(30,1.5)(0,0)
                     \multiput(0,0)(3,0){5}{\photonatomright}
                  \end{picture}
                 }

\def\photonup{\begin{picture}(1.5,30)(0,0)
                  \multiput(0,0)(0,3){10}{\photonatomup}
               \end{picture}
              }
\def\photonuphalf{\begin{picture}(1.5,15)(0,0)
                      \multiput(0,0)(0,3){5}{\photonatomup}
                   \end{picture}
                  }

\def\fermionup{\begin{picture}(1,30)(0,0)
                     \put(0,0){\vector(0,1){15}}
                     \put(0,15){\line(0,1){15}}
               \end{picture}
              }
\def\fermionuphalf{\begin{picture}(1,15)(0,0)
                         \put(0,0){\vector(0,1){7.5}}
                         \put(0,7.5){\line(0,1){7.5}}
                   \end{picture}
                  }

\def\fermionullhalf{\begin{picture}(15,7.5)(0,0)
                        \put(0,0){\vector(-2,1){7.5}}
                        \put(-7.5,3.75){\line(-2,1){7.5}}
                  \end{picture}
                 }

\def\fermionurrhalf{\begin{picture}(15,7.5)(0,0)
                        \put(-15,-7.5){\vector(2,1){7.5}}
                        \put(-7.5,-3.75){\line(2,1){7.5}}
                  \end{picture}
                 }

\newenvironment{Feynman}[3]{\begin{center}
                            \setlength{\unitlength}{#3 mm}
                            \begin{picture}(#1)(#2)
                            \thicklines
                           }{\end{picture} \end{center}}
\title{Initial state radiation to off-shell \zz~pair production in
       \ee annihilation\thanks{\noindent Contribution to the
                                $XVII^{th}$ Kazimierz Meeting
                                on Elementary Particle Physics,
                                Kazimierz, Poland, May
                                $23^{rd}-27^{th}$  1994.
                              }
      }
\author{Dietrich Lehner \\
        {\it DESY -- Institut f\"ur Hochenergiephysik} \\
        {\it D -- 15738 Zeuthen, Germany}
       }
\date{ }
\begin{document}
\maketitle
\begin{abstract}
\noindent
A study of the Standard Model reaction $\EE \rightarrow (\ZZ\ZZ)
\rightarrow
f_1\bar{f_1}f_2\bar{f_2}$ including the effects of the finite
\zz~width and initial state radiative corrections is presented. All
angular phase space integrations are
performed analytically. The remaining invariant masses are
integrated numerically. Semi-analytical and numerical
results in the energy range $\RS=200\;GeV$ to $1\;TeV$ are reported.
\end{abstract}
%
%
\section{Introduction}
%
For \ltwo~and a planned $500\;GeV$ \ee~collider, annihilation into
boson pairs is a major issue, because double resonance production is
strongly enhanced. \ltwo will operate above the \ww~pair production
threshold, perhaps above the $\ZZ\!\ZZ$~threshold and will open a new
discovery
window for a Higgs boson in the $\ZZ\!H$ channel. Since all heavy
bosons decay, their finite widths must be taken into account.
Furthermore radiative corrections are needed to match theoretical with
experimental precision. Thus, numerous efforts were made to
theoretically describe boson pair production and four fermion final
states.
\para
In this paper I restrict myself to the unpolarized reaction
\bq
  \EE \rightarrow (\ZZ\ZZ) \rightarrow f_1\bar{f_1}f_2\bar{f_2}~,
  ~~~~~~~~~~~~ f_1\!\neq\!f_2~,~~~f_i\!\neq\!e.
  \label{eezz4f}
\eq
including the effects of
the finite \zz~width and QED Initial State Radiation (ISR). The
motivation
for this is fourfold. Process~(\ref{eezz4f}) should be measurable at
\ltwo energies and above. It
represents a genuine higher order test of the electroweak Standard
Model, and, for some channels, it yields a background for \ww~physics as
well as Higgs boson searches. In addition, this study can be
easily extended to nonstandard neutral current physics, e.g.
$Z'$~bosons exchanges.
\para
On-shell \zz~pair production has been discussed long
ago~\cite{Brown78}. Numerical calculations including all
\oalf~electroweak corrections except hard photon brems\-strah\-lung
were reported in~\cite{Denner88} for on-shell and in~\cite{Denner90}
for off-shell \zz~bosons.
Monte Carlo generators including radiative corrections were described
in~\cite{FERMISV,Pittau94}.
On-shell calculations fail to take the important
finite width effects into account. Furthermore ISR is treated
incompletely in references~\cite{Denner88,Denner90}.
References~\cite{Denner90,FERMISV,Pittau94} strongly rely on numerical
phase space integration. In this note, in an effort to augment the
understanding of
process~(\ref{eezz4f}), off-shell complete ISR is calculated
following
a semi-analytical approach. This means that all angular phase space
variables are integrated analytically, leaving three invariant masses
for numerical integration. Quasi-experimental cuts on these can be
easily implemented.
\para
The paper is organized as follows. In section 2 semi-analytical
results for the off-shell Born cross-section are presented, followed
by the ISR results in section 3. Section 4 contains a short discussion
of backgrounds to process~(\ref{eezz4f}) and a few remarks on the
results' gauge behaviour. Section 5 closes the paper with a
summary, an outlook and conclusions.
%
%
\section{The Born Cross-Section}
\begin{figure*}
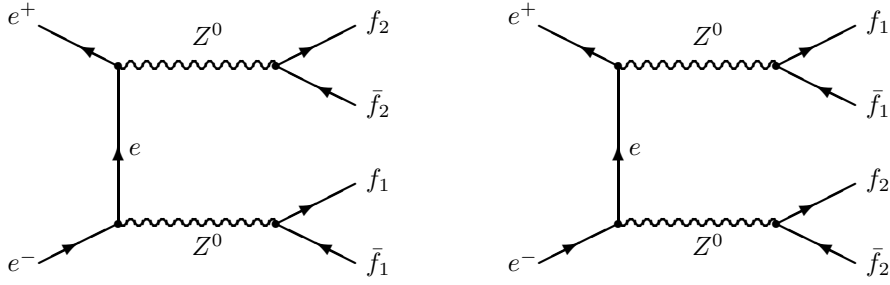

\vspace*{.2cm}
\begin{minipage}[tbh]{7.8cm}{
\begin{center}
\begin{Feynman}{75,60}{-20,0}{0.7}
%
\put(5,15){\fermionurrhalf}
\put(5,45){\fermionullhalf}
\put(5,15){\fermionup}
\put(5,45){\photonright}
\put(5,15){\photonright}
\put(5,45){\circle*{1.5}}
\put(5,15){\circle*{1.5}}
\put(35,15){\circle*{1.5}}
\put(35,45){\circle*{1.5}}
\put(50,7.5){\fermionullhalf}
\put(50,22.5){\fermionurrhalf}
\put(50,37.5){\fermionullhalf}
\put(50,52.5){\fermionurrhalf}
\small
\put(7,28){$e$}
\put(-16,6){\emi}
\put(-16,53){\epl}
\put(19,49){\zz}  
\put(19,09){\zz}
\put(52,06){${\bar f}_1$} 
\put(52,22){$ f_1$}
\put(52,36){${\bar f}_2$}
\put(52,52){$ f_2$}
\normalsize
\put(100,15){\fermionurrhalf}
\put(100,45){\fermionullhalf}
\put(100,15){\fermionup}
\put(100,45){\photonright}
\put(100,15){\photonright}
\put(100,45){\circle*{1.5}}
\put(100,15){\circle*{1.5}}
\put(130,15){\circle*{1.5}}
\put(130,45){\circle*{1.5}}
\put(145,7.5){\fermionullhalf}
\put(145,22.5){\fermionurrhalf}
\put(145,37.5){\fermionullhalf}
\put(145,52.5){\fermionurrhalf}
\small
\put(102,28){$e$}
\put(79,6){\emi}
\put(79,53){\epl}
\put(114,49){\zz}  
\put(114,09){\zz}
\put(147,06){${\bar f}_2$} 
\put(147,22){$ f_2$}
\put(147,36){${\bar f}_1$}
\put(147,52){$ f_1$}
\normalsize
\end{Feynman}
\end{center}
}\end{minipage}
\vspace{-.5cm}
\caption{\it The Born level Feynman diagrams for off-shell \zz~pair
         production. Left: t-channel. Right: u-channel.
        }
\label{zzfeyn}
\vspace{-.8cm}
\end{figure*}

At Born level, process~(\ref{eezz4f}) is described by the two Feynman
diagrams depicted in figure~\ref{zzfeyn}.
The Born cross-section is given by a simple convolution formula
\ba
\sigma^{ZZ}(s) & = &
\int\limits_{4m_1^2}^s ds_1 \, \rho_Z(s_1)
\int\limits_{4m_2^2}^{(\sqrt{s} - \sqrt{s_1})^2} ds_2 \, \rho_Z(s_2)
\cdot \sigma^{ZZ}_4(s;s_1,s_2)
\label{sigzz}
\ea
invoking Breit-Wigner density functions for the s-channel
\zz~propagators:
\ba
\rho_Z(s_i)
=
\frac{1}{\pi}
\frac {\sqrt{s_i} \, \Gamma_Z (s_i)
\times  B\!R(i)}
      {|s_i - M_Z^2 + i \sqrt{s_i} \, \Gamma_Z (s_i) |^2}.
\label{rhoz}
\ea
$B\!R(i)$ is the branching ratio for the decay channel under study,
$s_1$ and $s_2$ are the invariant \zz~masses. The \zz~width is given
by
\bq
\Gamma_Z (s_i) =
  \frac{G_{\mu}\, M_Z^2} {24\pi \sqrt{2}} \sqrt{s_i}
  \sum_f (v_f^2+a_f^2).
\label{gzoff}
\eq
\hspace*{-.25cm}
Using $a_e\!\!=\!\!1, \,v_e\!\!=\!\!1\!-\!4\sin^2 \!\theta_W,
\,L_e\!\!=\!\!(a_e\!+\!v_e)/2$ and $R_e\!\!=\!\!(a_e\!-\!v_e)/2\;$,
$\;\sigma^{ZZ}_4(s;s_1,s_2)$ is obtained after fivefold
analytical integration over the angular phase space variables.
\ba
\sigma^{ZZ}_4(s;s_1,s_2)
& = & {\displaystyle \frac{\left(G_{\mu} M_Z^2 \right)^2}{8\pi s}} \!
\left(L_e^4+R_e^4\right) {\cal G}_4^{t+u}(s;s_1,s_2).
\label{sigzz4}
\ea
The sub-index 4 indicates that the underlying matrix element contains
four resonant pro\-pa\-ga\-tors.
Although ${\cal G}_4^{t+u}$ is the sum of three kinematical
functions stemming from the t-channel, the u-channel, and the t-u
interference, it can be very compactly written as:
\ba
   {\cal G}_4^{t+u}(s;s_1,s_2)
   = \frac{\lambda^{1/2}}{s}\left[\frac{s^2+(s_1+s_2)^2}{s-s_1-s_2}
     {\cal L}_4 -2 \right]
\label{zzmuta}
\ea
with~ $\lambda \equiv s^2 + s_1^2 + s_2^2 - 2ss_1 - 2 s_1s_2 - 2 s_2s$
{}~~and
\bq
{\cal L}_4(s;s_1,s_2) =
\frac{1}{\sqrt{\lambda}} \, \ln \frac{s-s_1-s_2+\sqrt{\lambda}}
                                     {s-s_1-s_2-\sqrt{\lambda}}~.
\label{L4}
\eq
This result was only recently derived~\cite{teudb94}.
In the on-shell limit, eq.~(\ref{rhoz}) yields
$\rho_Z(s_i)=\delta(s_i - M_Z) \!\cdot \!B\!R(i)$ and,
with $\beta_Z\!=\!\sqrt{1-4M_Z^2/s\;}$, eq.~(\ref{zzmuta}) becomes
\bq
  {\cal G}_4^{t+u}(s;M_Z,M_Z) = 2 \cdot
  \left[\frac{1+4M_Z^2/s}{1-2M_Z^2/s}
        \ln \frac{1+\beta_Z}{1-\beta_Z} - \beta_Z
  \right] \; ,
\eq
being in agreement with~\cite{Brown78}. The effect of the finite
\zz~width can be seen from figure~\ref{zzxsec} as the characteristic
smearing of the peak.
\begin{figure}[h]
\vspace{11.5cm}
\vspace{-.85cm}
\caption[\zz total cross-section.]
{\it The total cross-section $\sigma^{ZZ}(s)$ for
  process~(\ref{eezz4f}).}
\vspace{-1cm}
\label{zzxsec}
\end{figure}
%
%
\section{${\cal O}(\alpha)$ Initial State Radiation}
\begin{figure}[b]
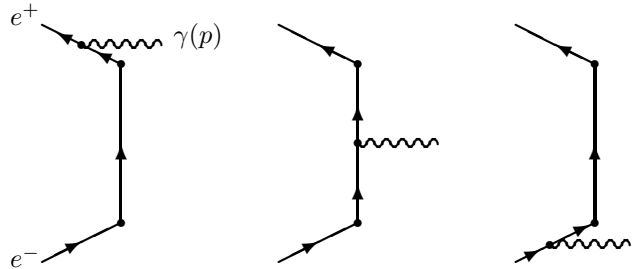

\begin{minipage}[tbh]{15.cm} {
\begin{center}
\begin{Feynman}{150,60}{-43,0}{0.7}
\small
\put(-26,52.5){$e^+$}  
\put(-26,06.5){$e^-$}
\put(5,49){$\gamma(p)$}
\normalsize
\put(-5,45){\line(-2,1){15.00}}
\put(-16.25,50.625){\vector(-2,1){1}}  
\put(-8.75,46.875){\vector(-2,1){1}}
\put(-5,15){\fermionurrhalf}
\put(-5,15){\fermionup}
\put(-12.5,48.75){\photonrighthalf}
\put(-5,45){\circle*{1.5}}
\put(-5,15){\circle*{1.5}}
\put(-12.5,48.75){\circle*{1.5}}
%
\put(85,45){\fermionullhalf}
\put(85,15){\fermionup}
\put(76.5,10.75){\photonrighthalf}
\put(85,45){\circle*{1.5}}
\put(85,15){\circle*{1.5}}
\put(76.5,10.75){\circle*{1.5}}
\put(70,7.5){\line(2,1){15.00}}
\put(73.75,9.375){\vector(2,1){1}}  
\put(83,14){\vector(2,1){1}}
%
\put(40,15){\fermionurrhalf}
\put(40,45){\fermionullhalf}
\put(40,30){\fermionuphalf}
\put(40,15){\fermionuphalf}
\put(40,30){\photonrighthalf}
\put(40,45){\circle*{1.5}}
\put(40,15){\circle*{1.5}}
\put(40,30){\circle*{1.5}}
\end{Feynman}
\end{center}
}\end{minipage}
\vspace{-.5cm}
\caption{\it The amputated ISB diagrams for \zz~pair production.}
\label{zzbrem}
\end{figure}

\begin{figure}[t]
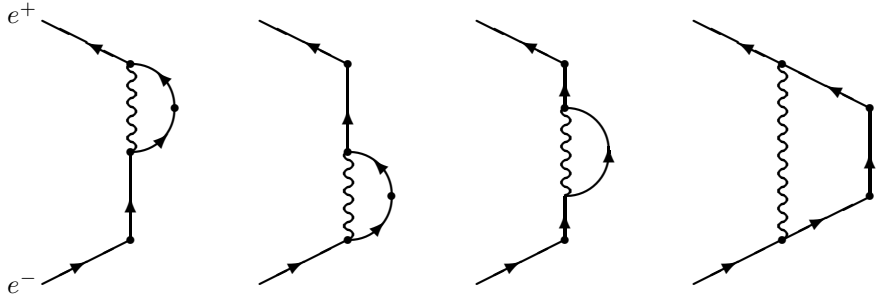

\vspace{.3cm}
\begin{minipage}[tbh]{15cm} {
\begin{center}
\begin{Feynman}{150,60}{-44,0}{0.78}
\small
\put(-42,52){$e^+$}  
\put(-42,6){$e^-$}
\put(-21,15){\fermionurrhalf}
\put(-21,45){\fermionullhalf}
\put(-21,15){\fermionuphalf}
\put(-21,30){\photonuphalf}
\put(-21,37.5){\oval(15,15)[r]}
\put(-15.6,42.8){\vector(-1,1){1}}  
\put(-15.55,32.2){\vector(1,1){1}}  
\put(-21,45){\circle*{1.5}}
\put(-21,15){\circle*{1.5}}
\put(-21,30){\circle*{1.5}}
\put(-13.5,37.5){\circle*{1.5}}
\put(16,15){\fermionurrhalf}
\put(16,45){\fermionullhalf}
\put(16,30){\fermionuphalf}
\put(16,15){\photonuphalf}
\put(16,22.5){\oval(15,15)[r]}
\put(21.35,27.8){\vector(-1,1){1}}  
\put(21.35,17.2){\vector(1,1){1}}  
\put(16,45){\circle*{1.5}}
\put(16,15){\circle*{1.5}}
\put(16,30){\circle*{1.5}}
\put(23.50,22.5){\circle*{1.5}}
\put(53,15){\fermionurrhalf}
\put(53,45){\fermionullhalf}
\put(53,15){\line(0,1){7.5}}
\put(53,18.75){\vector(0,1){1}}  
\put(53,41.25){\vector(0,1){1}}  
\put(53,37.5){\line(0,1){7.5}}
\put(53,22.5){\photonuphalf}
\put(53,30){\oval(15,15)[r]}
\put(60.48,30){\vector(0,1){1}}  
\put(53,45){\circle*{1.5}}
\put(53,15){\circle*{1.5}}
\put(53,37.5){\circle*{1.5}}
\put(90,15){\fermionurrhalf}
\put(105,22.5){\fermionurrhalf}
\put(90,45){\fermionullhalf}
\put(105,37.5){\fermionullhalf}
\put(105,22.5){\fermionuphalf}
\put(90,15){\photonup}
\put(90,45){\circle*{1.5}}
\put(90,15){\circle*{1.5}}
\put(105,22.5){\circle*{1.5}}
\put(105,37.5){\circle*{1.5}}
\end{Feynman}
\end{center}
}\end{minipage}
\vspace{-.6cm}
\caption{\it The amputated virtual ISR diagrams for \zz~pair
         production.}
\vspace{-.5cm}
\label{zzvirt}
\end{figure}

In \ee~annihilation, ISR is known to represent the bulk of the
radiative corrections.
The ${\cal O}(\alpha)$ `amputated' Feynman diagrams for Initial State
Bremsstrahlung (ISB) to process~(\ref{eezz4f})
are depicted in figure~\ref{zzbrem}. The corresponding
virtual ISR diagrams are shown in
figure~\ref{zzvirt}. External leg self energies are absorbed into the
on-shell renormalization.
The double-differential cross-section for off-shell \zz~pair
production including ${\cal O}(\alpha)$ ISR with soft photon
exponentiation can be presented as
\bq
  \frac{d^2 \sigma^{ZZ}}{ds_1 ds_2}
  =
  \int\limits_{(\sqrt{s_1}+\sqrt{s_2})^2}^s \!\!\!\!
  \frac{ds'}{s}
  \, \rho(s_1) \, \rho(s_2) \,
  \left[ \beta_e v^{\beta_e - 1} {\cal S} + {\cal H} \right]
  \label{compqed}
\eq
with $\beta_e \!=\!2 \alpha/\pi [ \ln (s/m_e^2) - 1 ]$ and $v=1-s'/s$.
The soft+virtual and hard photonic parts ${\cal S}$ and ${\cal H}$
are calculated analytically, requiring seven angular integrations.
Both separate into a universal part with the Born cross-section
factorizing and a nonuniversal part:
\ba
  {\cal S}(s,s';s_1,s_2) & = &
  \left[1 + {\bar S}(s) \right] \sigma_0(s';s_1,s_2)
  + \;\sigma_{\hat S}(s';s_1,s_2) ~~~~,
  \nl
  \hspace{-1cm}
  {\cal H}(s,s';s_1,s_2) & = &
  \underbrace{{\bar H}(s,s') ~\sigma_0(s';s_1,s_2)~~~}_{Universal\;
    Part}
  + \underbrace{\sigma_{\hat H}(s,s';s_1,s_2)}_{Nonuniversal\;Part}.
\ea
An explicit derivation proved that as expected ${\bar S}$ and
${\bar H}$ are identical to the radiators known
from s-channel fermion pair production~\cite{radiators}:
\ba
  {\bar S}(s) = \frac{\alpha}{\pi}
                \left[  \frac{\pi^2}{3} - \frac{1}{2} \right]
                + \frac{3}{4}\beta_e + {\cal O}(\alpha^2) ,
  \nl
  {\bar H}(s,s') = - \frac{1}{2}
                     \left(1+\frac{s'}{s}\right)\beta_e
                   + {\cal O}(\alpha^2).
\ea
The analytical calculation of the non\-uni\-ver\-sal
con\-tri\-bu\-tions is under way, its brems\-strah\-lung part already
completed. Since the corresponding analytical expressions contain
many Spence functions, nonuniversal contributions are involved. On the
other hand they are small, because
they do not contain the mass singularity $\beta_e$. Similar arguments
hold for \ww~pair production~\cite{teudb94,wwanal93}.
The off-shell cross-section of process~(\ref{eezz4f}), corrected
for universal ISR, is presented in figure~\ref{zzxsec}.
It is radiatively reduced below the peak and develops a
strong radiative tail above.
%
%
\section{Background}
%
\begin{figure}[bt]
\vspace{12cm}
\vspace{-2.3cm}
\caption[\zz total cross-section after cut.]
{\it Effect of a cut $|s_i-M_Z| \!\leq\! 10\;GeV$ on the off-shell
  Born cross-section.}
\label{zzxscut}
\vspace{-.5cm}
\end{figure}
\noindent
Strictly speaking, gauge invariance requires that all Feynman diagrams
contributing to $\EE \rightarrow f_1\bar{f_1}f_2\bar{f_2}$ be
taken into account. This means the inclusion of not only singly and
nonresonant diagrams as
given in figure~2 of~\cite{teudb94}, but also diagrams with photons
replacing the \zz~bosons and, for some final states, charged
current diagrams. However, as they are suppressed by a factor
$\Gamma_Z/M_Z$~for each nonresonating boson propagator, singly and
nonresonant diagrams only yield small
contributions~\cite{Pittau94,teup94bg}.
Their smallness for the case of
\ww~pair production was proven in~\cite{Pittau94,teudb94}.
Photon exchange
diagrams can also be considered as background, if, similar to
experimental situations, invariant mass cuts are applied to isolate
decaying \zz~bosons. Such cuts essentially extinguish photon exchange
contributions, but have a small effect on resonant diagrams as can be
seen from figure~\ref{zzxscut}.
\\
Slight gauge violations also come with the introduction of finite
boson widths. A scheme to avoid these was proposed, but
gives incorrect results around threshold~\cite{wwgauge}.
Up to now, no scheme for the introduction of finite
widths into boson pair processes seems theoretically satisfactory.
However, different schemes only differ in higher order.
%
%
\section{Summary, Outlook and Conclusions}
I have reported finite width and initial state QED corrections to
$\EE \!\!\rightarrow \!(\ZZ\ZZ) \!\rightarrow f_1\bar{f_1}f_2\bar{f_2}$ in a
semi-analytical approach. It was shown that both yield important
corrections to the total cross-section. Final State Radiation (FSR)
and Initial-Final Interferences (IFI) become important at very high
energies. FSR can be implemented in a radiator approach, but it is
unclear, if the semi-analytical treatment is efficient for IFI.
\para
Next, the calculation of nonuniversal contributions to
ISR has to be completed. After this the inclusion of photon exchange
graphs is natural. For the future it is planned to compute
singly and nonresonant background, its contribution to the
cross-section being of comparable magnitude as nonuniversal ISR.
Further
plans comprise FSR and IFI calculations. For the more remote future,
the inclusion of weak corrections and non-standard neutral current
physics are likely extensions of the presented work.
\para
I am obliged to the organizers of the ``$XVII^{th}$ Kazimierz
Meeting on Elementary Particle Physics'' for their hospitality, the
excellent organization, and the
stimulating atmosphere. Illuminating discussions with D. Bardin and T.
Riemann are gratefully acknowledged.
%
%

%
\end{document}